\documentstyle[prl,aps,multicol,psfig]{revtex}

\def\ul{ }

\begin{document}
\draft
\title {Nonequilibrium spin fluctuations in single-electron transistors}
\date{Received \hspace{5mm} }
\date{\today}

\author {Jan Martinek$^{1,2,3}$, J\'ozef Barna\'s$^{3,4}$,
Gerd Sch\"{o}n$^{2,5}$, Saburo Takahashi$^1$ and Sadamichi
Maekawa$^1$ }
\address{
 $^1$Institute for Materials Research, Tohoku University, Sendai
980-8577, Japan\\
 $^2$Institut f\"{u}r Theoretische
Festk\"{o}rperphysik,
 Universit\"{a}t Karlsruhe, 76128 Karlsruhe, Germany \\
 $^3$Institute of Molecular Physics, Polish
Academy of Sciences,
 ul. Smoluchowskiego 17, 60-179 Pozna\'n, Poland\\
 $^4$Department of Physics, Adam Mickiewicz
University, ul. Umultowska 85, 61-614 Pozna\'n, Poland\\
 $^5$Forschungszentrum Karlsruhe, Institut f\"ur Nanotechnologie,
76021 Karlsruhe, Germany \\
}
 \maketitle

\begin{abstract}
We show that nonequilibrium spin fluctuations significantly
influence the electronic transport in a single-electron
transistor, when the spin relaxation on the island is slow
compared to other relaxation processes, and when size effects play
a role. To describe spin fluctuations we generalize the `orthodox'
tunneling theory to take into account the electron spin, and show
that the transition between consecutive charge states can occur
via a high-spin state. This significantly modifies the shape of
Coulomb steps and gives rise to additional resonances at low
temperatures. Recently some of our predictions were confirmed by
Fujisawa et al. [Phys. Rev. Lett. {\bf 88}, 236802 (2002)], who
demonstrated experimentally the importance of nonequilibrum spin
fluctuations in transport through quantum dots.
\end{abstract}



\begin{multicols}{2}

Electronic transport in single electron transistors (SETs) with
metallic islands or semiconducting quantum dots (QDs) was usually
described in terms of the `orthodox' theory  of sequential
tunneling \cite{Sohn}.
The relaxation processes on the island were characterized only by
the energy relaxation time $\tau_{\rm \epsilon}$, with the tacit
assumption that the spin-flip relaxation time $\tau_{\rm sf}$ is
short or comparable to $\tau_{\rm \epsilon}$
\cite{Averin1,Beenakker,Aleiner}. In real systems, however, the
ratio $\tau_{\rm sf}/\tau_{\rm \epsilon}$ can be as high as $10^3$
- $10^4$ \cite{Khaetskii,Paillard,Fabian}. In this letter we show
that, when the spin relaxation time is significantly longer than
the energy relaxation time $\tau_{\rm \epsilon}$ and injection
time $\tau_{\rm I}$, large nonequilibrium spin fluctuations (NSF)
arise on the island. When the density of states for the island is
low, e.g., due to size effects, the spin fluctuations lead to
fluctuations of the spin-splitting of the chemical potential.
This, in turn, can significantly modify the transport
characteristics. In particular, the Coulomb steps are smeared
because transitions to higher charge states may occur via a set of
associated spin states, each of them corresponding to different
temporary electrochemical potentials. Such NSF can limit the
performance of spin electronic devices. The importance of spin
fluctuations has been recently confirmed experimentally by
Fujisawa {\sl et al.} \cite{Fujisawa} for a few-electron QD.

The effects described here differ from other phenomena related to
energy level quantization and electron spin, which have been
investigated recently theoretically \cite{Averin1,Beenakker} and
experimentally for paramagnetic \cite{Davidovic,Deshmukh,Delft}
and ferromagnetic grains \cite{Gueron,Kleff}. They also differ
from further spin-related phenomena, like spin blockade due to
exchange interaction \cite{Kramer}, parity effects \cite{Cobden},
Kondo effects \cite{Glazman,Gordon}, and quantum entanglement.

Before we analyze spin effects in normal metal SETs, we consider
first the corresponding double tunnel junction without Coulomb
blockade and with a linear current-voltage relation. Tunneling of
a single electron increases or decreases the magnetization $M$ of
the island by $1$. Here $M = N_{\uparrow} - N_{\downarrow} $,
where $ N_{\uparrow} \; ( N_{\downarrow}) $ are the numbers of
excess electrons with spin $ \sigma = \uparrow ( \downarrow )$. In
the absence of intrinsic spin-flip processes, the time evolution
of the magnetization $M(t)$ can be mapped onto a one-dimensional
diffusion process with $\langle M \rangle = 0$ and $ \langle M^2
\rangle^{1/2} \sim \sqrt{t}$. There are, however, two main
processes which restrict the increase of the fluctuations: (i)
spin-flip relaxation processes, and (ii) spin splitting of the
electrochemical potential $\mu_{\uparrow} \neq \mu_{\downarrow}$
(two spin components of the island can be treated as two
independent electron reservoirs with different electrochemical
potentials $\mu_{\sigma}$) due to spin accumulation $M$
\cite{accumulation}, which modifies the tunneling probability. In
the non-equilibrium situation due to an applied bias $V$,
fluctuating numbers $N_{\sigma}$ give rise to nonequilibrium
fluctuations $N_{\sigma} \Delta E$ of the electrochemical
potentials $\mu_{\sigma}$ (here $ \Delta E = 1/D(E_{\rm F})$ and
$D(E_{\rm F})$ denotes the density of states at the Fermi level in
the island), which influence the transport current. The currents
flowing through the left $L$ junction for both spin orientations
are $I_{\rm L \uparrow} = 1/2 \; (V R_{\rm L}/R + M \Delta E/2
e)/R_{\rm L}$ and $I_{\rm L \downarrow} = 1/2 \; (V R_{\rm L}/R -
M \Delta E/2 e)/R_{\rm L}$. $R_{\rm R}$ ($R_{\rm L}$) denote the
resistance of the right (left) junctions and $R=R_{\rm R}+R_{\rm
L}$. However, the total current $I_{\rm L} = I_{\rm L
\uparrow}+I_{\rm L \downarrow} = V/R$ is independent of $M$.
Straightforward calculation shows that for $eV/\Delta E \gg 1$ and
for low temperatures the magnetization fluctuations are described
by a Gaussian distribution with the standard deviation $ \langle
M^2 \rangle^{1/2} \approx \sqrt{(e V/\Delta E) \; a/(a+1)^2} $,
where $a = R_{\rm R}/R_{\rm L}$.

In contrast to the example discussed above, NSF in systems with
nonlinear current-voltage characteristics can influence the
average transport current. We will describe now how spin
fluctuations influence transport through a SET, where the charge
fluctuations are strongly suppressed due to the Coulomb
interaction, but spin fluctuations can be relatively large.
For charging effects the relevant charging energy is $E_{\rm
C}=e^2/2C$, where $C = C_{\rm L} + C_{\rm R} + C_{\rm g}$ is the
total capacitance of the island, which is the sum of capacitances
of the left ($C_{\rm L}$) and right ($C_{\rm R}$) junctions and of
the gate ($C_{\rm g}$). The electronic transport through the
system in a stationary state is governed by the solution of the
generalized master equation \cite{Martinek}
\begin{eqnarray}\label{eq:1}
&&0 \! = \! \! - \! \left\{ \Gamma ( N_{\uparrow}, \!
N_{\downarrow} ) \! + \! \Omega_{\uparrow , \!  \downarrow}
(N_{\uparrow} , \! N_{\downarrow}) \! + \! \Omega_{\downarrow,
\uparrow} (N_{\uparrow} , \!  N_{\downarrow}) \right\} \!
 P(N_{\uparrow}, \!  N_{\downarrow} ) \nonumber \\
&&+  \Gamma_{\uparrow}^{+} \! (N_{\uparrow} \! - \!\! 1, \!
N_{\downarrow}) P( N_{\uparrow} \! - \!\! 1, \! N_{\downarrow})
\! + \! \Gamma_{\downarrow}^{+} \! ( N_{\uparrow}, \!
N_{\downarrow} \! - \!\! 1) P( N_{\uparrow}, \! N_{\downarrow} \!
- \!\! 1) \nonumber
\\
&&+ \Gamma_{\uparrow}^{-} \! (N_{\uparrow} \! + \!\! 1, \!
N_{\downarrow}) P( N_{\uparrow} \! + \!\! 1, \! N_{\downarrow})
\! + \! \Gamma_{\downarrow}^{-} \! ( N_{\uparrow}, \!
N_{\downarrow} \! + \!\! 1) P( N_{\uparrow}, \! N_{\downarrow} \!
+ \!\! 1) \nonumber
\\
&&+ \Omega_{\uparrow,\downarrow}(N_{\uparrow} - 1, N_{\downarrow}
+ 1) P( N_{\uparrow} - 1, N_{\downarrow} + 1 ) \nonumber \\
&&+ \Omega_{\downarrow,\uparrow}( N_{\uparrow} + 1, N_{\downarrow}
- 1) P( N_{\uparrow} + 1, N_{\downarrow} - 1)
\end{eqnarray}
for the probability $P(N_{\uparrow},N_{\downarrow})$ to find
$N_{\uparrow}$ and $N_{\downarrow}$ excess electrons on the island
($ N = N_{\uparrow} + N_{\downarrow}$ is the total number of
excess electrons). The first term in Eq.~(\ref{eq:1}) describes
the rate at which the probability of a given configuration decays
due to electron tunneling to or from the island, whereas other
terms describe the rate at which this probability increases. The
$\Omega$ terms account for spin-flip relaxation processes.
The coefficients entering Eq.~(\ref{eq:1}) are defined as
$\Gamma_{\sigma}^{\pm} (N_{\uparrow},N_{\downarrow}) = \sum_{r=
\rm L,R} \Gamma_{r \sigma}^{\pm} (N_{\uparrow},N_{\downarrow}) $
and
  $ \Gamma ( N_{\uparrow},
N_{\downarrow} ) =
 \sum_{p=\pm} \sum_{\sigma}
\Gamma_{\sigma}^p (N_{\uparrow},N_{\downarrow}) $,
where $\Gamma_{r\sigma}^{\pm} (N_{\uparrow},N_{\downarrow})$ are
the tunneling rates for electrons with spin $\sigma$, tunneling to
($+$) the grain from the lead ${r = \rm L,R}$ or back ($-$). These
coefficients are given by the following expression:
\begin{eqnarray}\label{2}
&& \Gamma^{\pm}_{r \sigma} \! (N_{\uparrow} ,\! N_{\downarrow}) \!
= \!\!\! \sum_{i} \! \gamma^r_{i \sigma }F^{\mp}_{\sigma} \!
(E_{i\sigma} \! | N_{\uparrow} , \! N_{\downarrow} ) f^{\pm} \!
(E_{i\sigma} \!\! + \!\! E^{+}_r(N) \!\! - \!\! E_F) ,  \nonumber
\\
&& \Omega_{\sigma\overline{\sigma}}(N_{\uparrow}, \!
N_{\downarrow} \! ) \!  = \!\! \sum_{i} \! \sum_{j} \!
\omega_{i{\sigma},j{\overline{\sigma}}}
 F^{+}_{\sigma} \! (E_{i\sigma} |
N_{\uparrow} , \! N_{\downarrow} )
 \nonumber \\
&& \hspace{3cm}  \times \,  F^{-}_{\overline{\sigma}} \!
(E_{j\overline{\sigma}} | N_{\uparrow} , \! N_{\downarrow} ) .
\end{eqnarray}
Here, $f^{+}(E)$ ($f^{-}=1-f^{+}$) is the Fermi distribution
function, whereas $ F^{+}_{\sigma}(E_{i\sigma} | N_{\uparrow}
,N_{\downarrow}) $ \cite{distribution}
($F^{-}_{\sigma}=1-F^{+}_{\sigma}$) describes the probability that
the energy level $E_{i\sigma}$ is occupied by an electron with
spin $\sigma$ for the particular configuration  $ (N_{\uparrow}
,N_{\downarrow})$.
The parameter $\gamma^r_{i\sigma}$ is the tunneling rate of
electrons between the lead $r$ and the energy level $E_{i\sigma}$
of the island, and $\omega_{i{\sigma},j{\overline{\sigma}}}$ is
the transition probability from the state $ i{\sigma} $ to
$j{\overline{\sigma}}$ due to the spin-flip processes.
The energies $E^{\pm}_{\rm L} (N)$ and $E^{\pm}_{\rm R} (N)$ are
given by $ E^{\pm}_{\rm L} (N)=C_{\rm R}/C \; e V +U^{\pm}(N)$ and
$ E^{\pm}_{\rm R} (N)=-C_{\rm L}/C \; e V +U^{\pm}(N)$ where
$U^{\pm}(N)= E_{\rm C} [ 2(N-N_x) \pm 1 ] $ and $N_x = C_{\rm g}
V_{\rm g}/e $, with $V_{\rm g}$ denoting the gate voltage.

From the solution $P(N_{\uparrow},N_{\downarrow})$ of the master
equation (1) we obtain the current flowing through the island
\begin{eqnarray}\label{eq:3}
I_r \! = \! e \! \! \sum_{\sigma} \! \! \! \sum_{\; N_{\uparrow}
,N_{\downarrow}} \! \! \! P(N_{\uparrow} ,N_{\downarrow}) \!
\left\{ \Gamma^{+}_{r \sigma}(N_{\uparrow} ,N_{\downarrow}) \! -
\! \Gamma^{-}_{r \sigma}(N_{\uparrow} ,N_{\downarrow})
 \right\} . \!
\end{eqnarray}
In our calculations we assume that discrete energy levels
$E_{i\sigma}$ are equally separated, with the level spacing
$\Delta E$. The tunneling rates $\gamma^r_{i\sigma}$ are then
given by the formula $ \gamma^r_{i\sigma} = \Delta E / e^2
R_{r\sigma}$ \cite{Beenakker}, where $R_{r\sigma}$ is the
resistance of the junction $r$ for spin $\sigma$ ($R_{r\uparrow}
=R_{r\downarrow}=2R_{r}$, in the case under consideration). For
spin-flip processes we assume
$\omega_{i{\sigma},j{\overline{\sigma}}} = 1/\tau_{sf} \;
\delta_{i,j}$.

\begin{figure}
\centerline{\psfig{figure=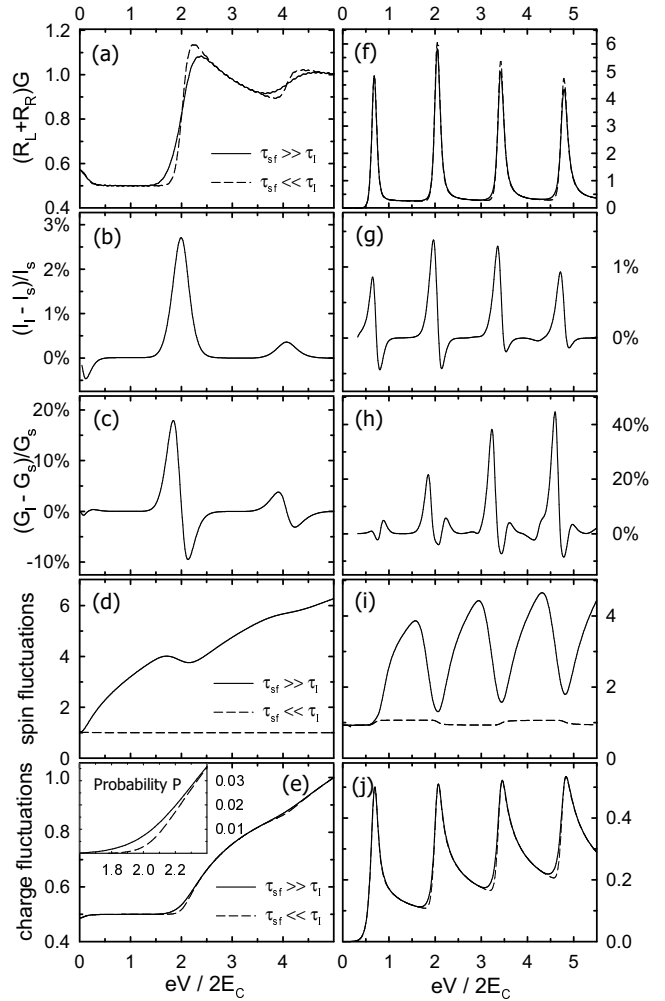,width=8.5 cm}} \caption{(a,f)
Differential conductance vs. bias voltage $V$ for two limits:
short (dashed line) and long (solid line) spin-flip relaxation
time; (b,g) relative difference $(I_{\rm l}-I_{\rm s})/I_{\rm s}$,
where $I_{\rm l}$ and $I_{\rm s}$ are the currents for the two
limits; (c,h) relative difference $(G_{\rm l} - G_{\rm s})/G_{\rm
s}$; (d,i) spin fluctuations $ \langle M^2 \rangle^{1/2} $; (e,j)
charge fluctuations $(\langle N^2 \rangle - \langle N
\rangle^2)^{1/2}$. The curves are calculated for symmetric (left
column) and asymmetric  junctions (right column). { \ul Inset: The
probability $P$ of the $N = 1$ charge state for a symmetrical
junction as a function of the bias voltage $V$. }
 For symmetric
junctions $C_{\rm R} = C_{\rm L}$, $R_{\rm R} = R_{\rm L}$ and the
gate voltage $N_x = 0.5 + 1/4 \; \Delta E / E_{\rm C}$ (at
resonance), whereas for asymmetrical junctions $C_{\rm R} / C_{\rm
L} = 3$, $R_{\rm R} / R_{\rm L} = 100$ and the gate voltage $N_x =
0$. The other parameters are $k_{\rm B}T = 0.4 \Delta E$ and
$\Delta E/E_{\rm C} = 0.1$ for both cases.
   }
\label{fig1}
\end{figure}

In Fig.~\ref{fig1} we show calculated transport characteristics
for symmetric (left column) and asymmetric (right column)
junctions. In both cases the transport characteristics are
calculated in the fast ($ \tau_{\rm I} \gg \tau_{\rm sf} , \tau_{
\rm \epsilon}$) and slow ($\tau_{\rm sf} \gg \tau_{\rm I} \gg
\tau_{\rm \epsilon}$) spin-flip relaxation limits, where $
\tau_{\rm I} \sim 1/\gamma^r_{i\sigma} \sim {\rm e}/I $ (for QDs,
it is shown in Ref.~\cite{Paillard} that $\tau_{\rm sf} > 1
\mu{\rm s} $, $\tau_{ \rm \epsilon} < 1{\rm ns} $ and $\tau_{\rm
I} \sim 1 \div 10 {\rm ns} $ for $I \sim 0.1 \div 1 {\rm nA} $).
In Fig.~\ref{fig1}(a,f) the conductance-voltage characteristics
are presented for both limits.
 The effect on the current is relatively
weak (a few per-cent at maximum), as shown in
Fig.~\ref{fig1}(b,g). However, the effect on the differential
conductance can be significantly larger, about a few tens per-cent
at maximum, as shown in Fig.~\ref{fig1}(c,h). Generally, the most
pronounced effect of the spin fluctuations is on the nonlinear
parts of the transport characteristics. This effect is similar to
the one produced by an increase in effective temperature $T_{\rm
eff}$ \cite{Korotkov,Garcia} of the system (different from the
bath temperature $T$). {\ul However, both effects can be easily
distinguished because of the parabolic dependence on the bias
voltage $V$ of the former effect (see the discussion below). For
intermediate spin-flip times, $0.1 < \tau_{\rm sf}/\tau_{\rm I} <
10$, the conductance smoothly crosses over from one limit to the
other.}

In Fig.~\ref{fig1}(d,i) we show the bias dependence of the spin
fluctuations $\langle M^2 \rangle^{1/2}$ in the limits of both
short and long spin relaxation times. In the former case the
fluctuations are almost constant and close to 1. In the latter
case and for symmetrical junctions, the fluctuations nearly follow
the law for junctions without Coulomb effects ($ \langle M^2
\rangle^{1/2} \approx \sqrt{(e V/\Delta E) \; a/(a+1)^2} $).
However, for asymmetric junctions there are pronounced
oscillations in the fluctuations amplitude with increasing bias
$V$, which are related to the charge accumulation and correlated
with the charge fluctuations \cite{Martinek1} (see
Fig.~\ref{fig1}(j)). For comparison, we show in
Fig.~\ref{fig1}(e,j) the amplitude of charge fluctuations. For
asymmetric junctions the oscillations are due to charge
accumulation.

In the inset of Fig.~\ref{fig1}(e), we show the probability $P$ of
the $N = 1$ state in the transition range (from $N =0$ to $N = 1$)
for the symmetrical junction.
 One can see that NSF in the case of slow spin-flip relaxation allow the
transition to the next charge state at lower bias voltage.

\begin{figure}
\centerline{\psfig{figure=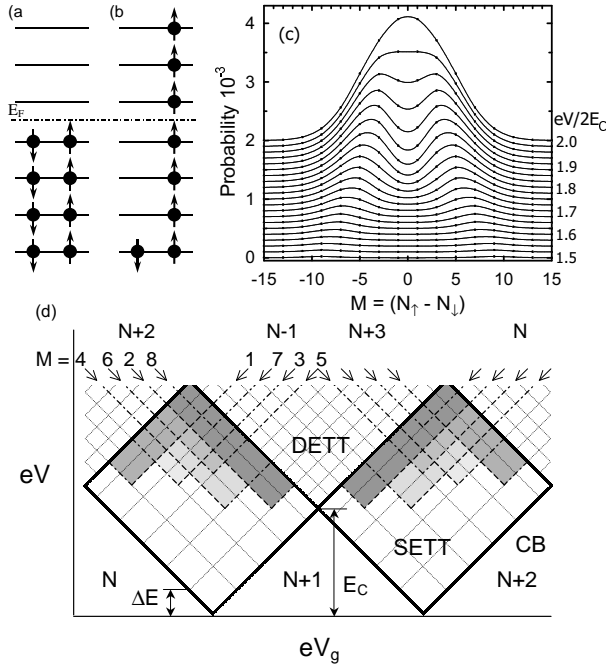,width=8 cm}}
 \caption{Examples of specific spin configurations of the island
for the same charge state  of the system with (a) short and (b)
long spin-flip relaxation times. Part (c) shows how the system
enters the next charge state. The probability $P(N_{\uparrow},
N_{\downarrow})$ for $N = (N_{\uparrow} + N_{\downarrow}) = 1$ is
shown there as a function of $ M = (N_{\uparrow} - N_{\downarrow})
$ for several values of the bias voltage $V$. All parameters are
the same as in Fig.~\ref{fig1}(a). The lines are a guide to eye
only. Data for different $V$ are offset vertically.
 {\ul (d) The
scheme of conductance peaks (current $I$ plateaus) for dc
excitation transport in the $V-V_{\rm g}$ plane for symmetric
junction, $E_{\rm C} / \Delta E = 4.5$  and $k_{\rm B}T \ll \Delta
E$. Solid lines determine the Coulomb blockade (CB) diamond
structure indicating also the onset of single electron tunneling
transport (SETT) and double electron tunneling transport (DETT)
(two excess electrons on the island are possible) in the absence
of NSF. Dotted lines are usual effects related to offset in
transport of the next discrete energy levels. Dashed and
dotted-dashed lines indicate the onset
 of the consecutive charge states due to particular spin
 excitations (as indicated) which without spin fluctuations are
 not accessible. Dashed lines indicate also formation of new
resonances due to NSF. Symmetry is broken due to the parity
effect.
 }
 }
 \label{fig3}
\end{figure}

\begin{figure}
\centerline{\psfig{figure=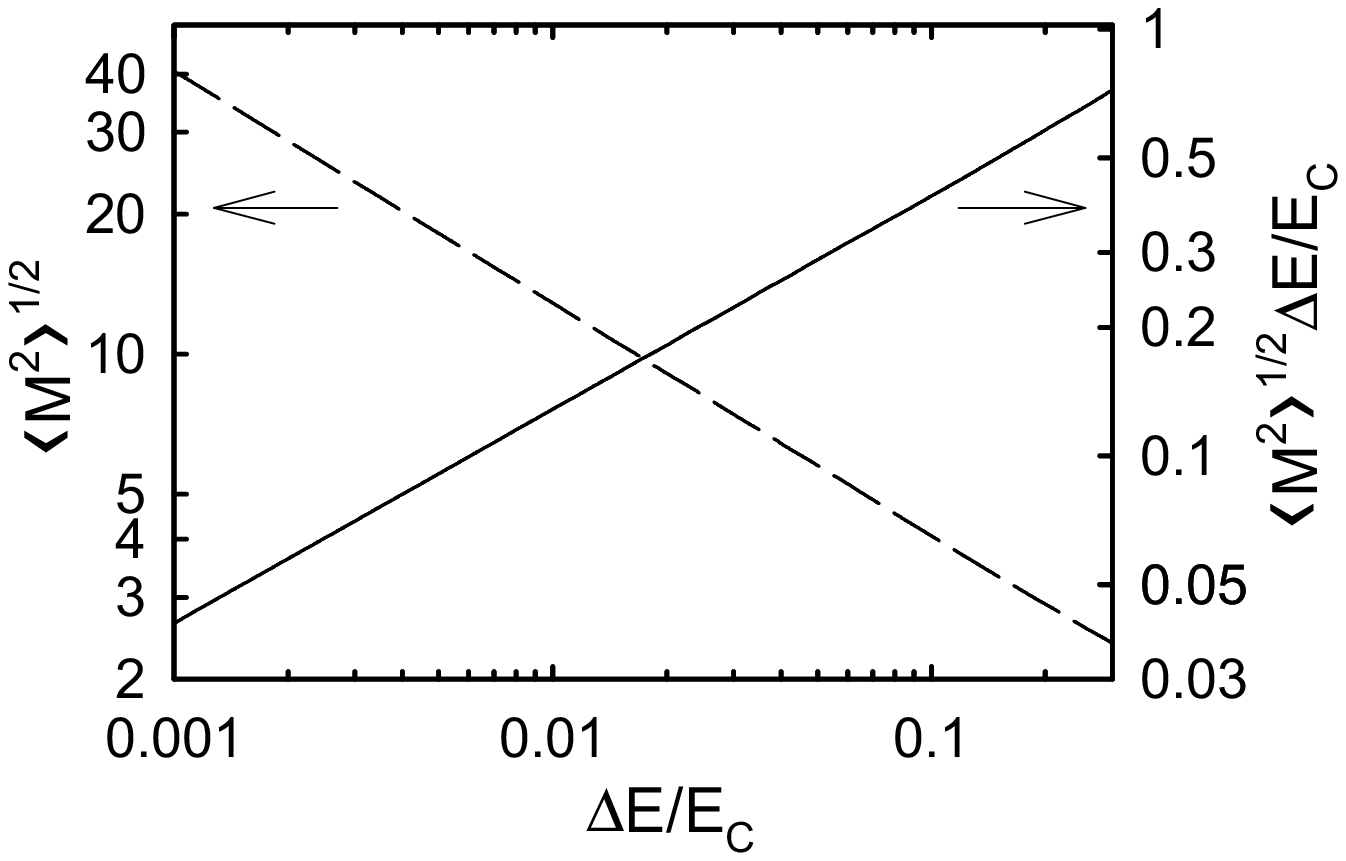,width=6.5 cm}} \caption{Spin
fluctuations $ \langle M^2 \rangle^{1/2} $ (dashed line) and
 the corresponding splitting of electrochemical potential $ \langle M^2
\rangle^{1/2} \Delta E / E_{\rm C}$ (solid line) as a function of
energy level spacing $\Delta E / E_{\rm C}$, calculated for
 $k_{\rm B}T = 0.04 E_{\rm C}$ and $ e V / 2 E_{\rm C} = 3 $. The other parameters
 are as in Fig.~\ref{fig1}(a). }
 \label{fig4}
\end{figure}

Figure~\ref{fig3} demonstrates the mechanism how the spin
fluctuations assist the system to enter the higher charge states.
For strong  NSF, an electron needs a lower energy to enter the
island, because some of the double degenerate states below the
Fermi energy are empty (compare Fig.~\ref{fig3}(b) and
Fig.~\ref{fig3}(a)). Figure~\ref{fig3}(c) illustrates that the
onset of the transition to the next charge state is linked to the
existence of a high spin state, and the new charge state at the
onset can appear only via the high spin state. The effect is
relevant as long as the energy of thermal fluctuations is lower
than the energy of NSF, $2\langle M^2 \rangle^{1/2} \Delta E >
k_{\rm B} T$. Thus, even if the discrete states are not resolved
($k_{\rm B} T \gtrsim \Delta E$), the effects due to spin
fluctuations can be important.
{\ul In Fig.~\ref{fig4} we show the amplitude of NSF as a function
of $\Delta E$. From this figure follows that the NSF effects can
be observed in electronic transport when $ \Delta E / E_{\rm C}
\gtrsim 0.01$. This condition can be easily achieved for typical
QDs \cite{Sohn} of radius $200 {\rm nm}$, $2 E_{\rm C} \sim 1 {\rm
meV}$, $\Delta E \sim 0.03 {\rm meV}$ and for $T < 1 K$. For
smaller QDs the effect is important also at higher temperatures.}
For larger islands with high density of states $D(E_{\rm F})$ the
fluctuations of $M$ have very large amplitude ($\sim
\sqrt{eV/\Delta E}$), but the associated effective spin splitting
of the electrochemical potential $M \Delta E$ is small ($\sim
\sqrt{eV \Delta E}$).
For metallic systems the condition $2\langle M^2 \rangle^{1/2}
\Delta E > k_{\rm B} T$ could be fulfilled for small metallic
particles with diameter in the range of several $nm$. For strong
exchange interactions QDs can have a ground state with spin $S >
1/2$ \cite{Kramer,Kurland}. But even in this case spin
fluctuations are important, as pointed out by Kleff and von Delft
\cite{Kleff} for ferromagnetic grains.

The time $\tau_{\rm sf}$  depends mainly on the strength of the
spin-orbit interaction and on the impurity contents \cite{Fabian}.
Thus counter-intuitively, for SETs with dirty island and strong
spin scattering one should expect sharper transport
characteristics.
 {\ul The NSF effects can be detected by measurements of the
conductance step or peak widths as a function of temperature. The
widths should saturate at a value that scales with the bias
voltage as $V^{1/2}$. In the absence of NSF there will be no
saturation observed, or the saturation will be at a value (related
to some $T_{\rm eff}$) which does not depend on the bias as
$V^{1/2}$.}

{\ul Very recently, Fujisawa {\sl et al.} \cite{Fujisawa} obtained
results for very small QDs at low temperatures, which confirm the
importance of NSF and which can be accounted for by our model. In
Fig.~\ref{fig3}(d) we show schematically the positions of the
conductance peaks (current steps), i.e., dc excitations in the
low-temperature limit, $k_{\rm B}T \ll \Delta E$, where discrete
energy levels are resolved. In addition to the standard resonance
peaks (corresponding to the dotted lines), there are new ones
indicated by the dashed lines which start inside the diamond
corresponding to single electron tunneling transport (SETT) (not
at its border) when a particular spin excitation appears. Double
electron tunneling transport (DETT) can occur already within the
SETT diamond region due to NSF (the regions marked in gray in
Fig.~\ref{fig3}(d), where also the magnetic state of the
excitations is indicated). We find that not only in results of
Ref.~\cite{Fujisawa} but also in earlier data (see e.g. Fig.1.(A)
in Ref.~\cite{Stewart}) the borders between SETT and DETT diamonds
could be affected by spin fluctuation. The new resonance peaks are
related to the fact that the ratio $E_{\rm C} / \Delta E$ is
generally not integer, and the electron levels for adjacent charge
states are effectively shifted by $mod(E_{\rm C}, \Delta E)$. }

One may expect a similar effect of NSF on the cotunneling current,
too. The fluctuations can influence in particular the first step
in the current-voltage characteristic (the first peak in the
conductance). One may also expect that NSF modify the current shot
noise.

In summary we have generalized earlier spinless descriptions of
SETs by taking into account the fact that spin-flip relaxation
time $\tau_{\rm sf}$ is usually much longer than the energy
relaxation time $\tau_{\rm \epsilon}$. The spinless models
describe properly only the situation when $\tau_{\rm sf}$ is short
or comparable to $\tau_{\epsilon }$. When, however, $\tau_{\rm sf}
\gg \tau_{\rm \epsilon}$, large nonequilibrium spin fluctuations
may occur in the island. These fluctuations have a significant
influence on the transport characteristics. Our model accounts for
recent experimental observations \cite{Fujisawa}.

This work is supported by a Grand-in-Aid for Scientific Research
from MEXT of Japan and the Polish State Committee for Scientific
Research under the Project No. 5 P03B 091 20. J.M. acknowledges
the hospitality of the University and Forschungszentrum Karlsruhe
and S.M. acknowledges the support of the Humboldt Foundation.


\end{multicols}


\begin{references}
\bibitem{distribution}
For $\tau_{\rm \epsilon} \ll \tau_{\rm I}$ (which is fulfilled in
this work), the distribution function $F_\sigma$ takes its
equilibrium form {\ul and can be calculated from the free energy
of internal degrees of freedom (for details see
Ref.~\cite{Martinek}) }. Thus, in the limits of $k_{\rm B}T \gg
\Delta E$ and $k_{\rm B}T \ll \Delta E$, the function $F_\sigma$
can be approximated by the Fermi distribution, but in the former
case the temperature $T$ should be replaced by $T/2$ (see
Ref.~\cite{Beenakker}). In other cases numerical calculations are
required.
\end{references}
\end{document}